\documentclass[11pt]{article}
\usepackage{graphics}
\usepackage{feynmf}
\usepackage{amsmath}
\begin{document}

\title{
  \textbf{Requirements to Detect the Monoenergetic}
  \textbf{Photon Signature of Thermal} 
  \textbf{Cold Dark Matter in PeV-Scale}
  \textbf{Split Supersymmetry}}
\author{Brooks Thomas \\ \\
\it{Michigan Center for Theoretical Physics (MCTP)} \\
\it{Department of Physics, University of Michigan, Ann Arbor, MI 48109}}
\date{March 23, 2005}
\maketitle

\begin{abstract}
Split supersymmetry scenarios with PeV-scale scalar masses circumvent many of the 
restrictions on supersymmetry coming from experimental limits on CP violation, 
flavor-changing neutral currents, and the Higgs boson mass.  We consider the
situation where the LSP is either a Wino or Higgsino
and the majority of its cold dark matter relic density is of thermal origin, in 
which case its 
mass is specified to be 2.3 TeV for a Wino or 1.1 TeV for a Higgsino.  We examine 
the potential for indirect detection, at present and
future \(\gamma\)-ray telescopes, of the monoenergetic photon lines 
that would result from the
annihilation of these particles near the galactic center.  
We show how the possibility for detection depends on the 
precise form of the galactic dark-matter halo profile and examine what performance 
attributes  a \(\gamma\)-ray detector would need in order to register 
a \(5\sigma\) discovery.   
\end{abstract}

\section{Theory}

\indent

If we ignore philosophical arguments regarding `naturalness', 
we can seriously consider a broader array of 
theories as being `physical' in the sense that they could realistically
describe our universe.  Such a reconsideration is both theoretically 
plausible, in light of string/M-theory landscape considerations, and
observationally motivated.  Because of data pressures on supersymmetric 
theories arising from Higgs mass
searches, limits on CP violation effects, etc., it would be advantageous if
one could assign to all the scalars in the theory (with the exception of one
light Higgs particle) masses at~\cite{Wells:2003tf} or 
above~\cite{Arkani-Hamed:2004fb,Giudice:2004tc} 
the PeV scale while keeping the 
gauginos relatively light, with TeV-scale masses, to 
constitute dark matter.  Taking anomaly-mediation as the origin of gaugino masses,
but not scalar masses~\cite{Randall:1998uk}, provides us with 
the desired mass hierarchy.  If the chiral supermultiplet \(S\) is charged under 
some symmetry, the scalar masses are generated via

\begin{equation}
  \mathcal{L}\supset
  \int d^{2}\theta d^{2}\overline{\theta}\frac{S^{\dagger}S}{M_{P}^{2}}
  \Phi_{i}^{\dagger}\Phi_{i}
  \longrightarrow \frac{F_{S}^{\dagger}F_{S}}{M_{P}^{2}}\phi_{i}^{\ast}\phi_{i},
\end{equation}
as usual, but the equation that would normally generate the gaugino masses  

\begin{equation}
  \mathcal{L}\supset
  \int d^{2}\theta\frac{S}{M_{P}^{2}}W^{a}W_{a}
  \longrightarrow \frac{F_{S}}{M_{P}^{2}}\lambda^{a}\lambda_{a}
\end{equation}
is no longer gauge invariant.  Instead, the leading contribution to the
gaugino masses arises at the one-loop level~\cite{Randall:1998uk} and takes
the form

\begin{equation}
  M_{\lambda}=\frac{\beta_{g_{\lambda}}}{g_{\lambda}}
  \frac{\langle F_{S}^{\dagger}F_{S}\rangle}{M_{P}^{2}},
\end{equation}         
where the index \(\lambda\) labels the three Standard Model gauge groups.
Explicit dependence of each gaugino mass on the beta function of the gauge
group with which it is associated determines the ratio between them, which
to lowest order is

\begin{equation}
  \label{eq:M123}
  M_{3} \simeq 3 M_{1} \simeq 9 M_{2}.
\end{equation}
If indeed the gravitino mass 
\(m_{3/2}^{2}=\langle F_{S}^{\dagger}F_{S}\rangle/M_{P}^{2}\) is at the PeV-scale 
all the scalar sparticles acquire PeV-scale masses while the gaugino masses are
kept light, around the TeV scale, in this construction, an example of what has
come to be known as 
``split supersymmetry''~\cite{Arkani-Hamed:2004fb,Giudice:2004tc,Arvanitaki:2004eu}.
Since the gaugino mass
hierarchy is determined by equation (\ref{eq:M123}), the lightest supersymmetric
particle must in general be some mixture of Wino and Higgsino (in more
general split supersymmetry models, the LSP may also have a significant Bino
fraction: this case is examined in~\cite{Pierce:2004mk}).  The only 
parameter yet unspecified in this scenario that has any bearing on the identity 
of the LSP is the \(\mu\) term.  If \(\mu>M_{2}\), the LSP is Wino-like; if 
\(\mu<M_{2}\), it is Higgsino-like; and if the two are comparable, it will be a 
mixture of the two.  

\indent
In this work, we examine the case in which the LSP is effectively either pure Wino 
or pure Higgsino, noting that if \(m_{\mathrm{LSP}}\gg m_{Z}\), these
two possibilities cover the vast majority of (\(M_{2}\), \(\mu\)) parameter space.  
If the Wino is the LSP, its thermal relic 
abundance~\cite{Giudice:2004tc} is given by

\begin{equation}
  \Omega_{\tilde{W}}^{therm}\simeq 0.02
  \left(\frac{M_{2}}{1 \mathrm{TeV}}\right)^2.
\end{equation}
For a Higgsino LSP, the result is

\begin{equation}
  \Omega_{\tilde{H}}^{therm}\simeq 0.09
  \left(\frac{\mu}{1 \mathrm{TeV}}\right)^2.
\end{equation}
We can use the current WMAP bound~\cite{Spergel:2003cb} on the abundance of cold 
dark matter in the universe

\begin{equation}
  \Omega_{\mathrm{CDM}}h^{2}=0.11\pm0.01 \mbox{ (WMAP 68\% C.L.)}
\end{equation}  
to find the mass the LSP would require in order to have an 
interesting relic density.  In this paper, we will make the simplifying assumptions 
that this particle makes up all the dark matter in the universe and that its relic 
density is generated thermally. the result, in each case, is

\begin{eqnarray}
  M_{\tilde{W}}\simeq2.3\pm0.2\mbox{ TeV} \label{eq:mw} \\
  M_{\tilde{H}}\simeq1.1\pm0.1\mbox{ TeV}.\label{eq:mh}
\end{eqnarray} 
If we allow for the possibility of other dark matter constituents, or for nonthermal
LSP generation, these equations become an upper bound.  In the purely thermal
case, it should be noted that in order to make up 
even 20\% of the dark matter in the universe, \( M_{\tilde{W}}\) would still 
need to exceed 1 TeV (\( M_{\tilde{H}}\) would be around 500 GeV)  
and that none of the particle properties relevant to the indirect detection 
methods discussed below (e.g. annihilation cross-section into \(\gamma\)-rays) 
varies significantly over this mass range.  If there are nonthermal
contributions to the LSP relic density, however, any value of \(M_{\mathrm{LSP}}\) 
below the bound given by equations (\ref{eq:mw}) and (\ref{eq:mh}) is permitted.

\indent 
Since theory has stipulated a TeV-scale Wino or Higgsino as our dark matter 
candidate, it is reasonable to ask whether any signature of such a particle 
could be detected experimentally.  Direct detection of such a massive Wino
or Higgsino LSP is effectively ruled out for all but the most unnaturally
peaked halo models\footnote{A thorough exposition of LSP dark matter constraints from
various planned and operational experiments, especially in mixed Wino-Higgsino
and Bino-Wino scenarios, can be found in~\cite{Masiero:2004ft}.} and the positron 
signal resulting 
from their annihilation would be too small to detect.  The only truly promising 
detection method is to search for high-energy photons produced monoenergetically 
by Wino annihilation in the galactic halo.  These photons would have energies of

\begin{eqnarray}
  E_{\gamma}=m_{\chi} & \hspace{1cm} \mathrm{and} \hspace{1cm} & 
  E_{\gamma}=m_{\chi}\left(1-\frac{m_{Z}^{2}}{4m_{\chi}^{2}}\right),
\end{eqnarray}
(corresponding to \(\chi\chi\rightarrow \gamma\gamma\) and 
\(\chi\chi\rightarrow \gamma Z\) processes respectively).  
Presumably, as the concentration of dark matter is greatest in the center of 
the galaxy, one could aim a sufficiently powerful \(\gamma\)-ray telescope 
at the galactic center and see a signal at around 2.3 TeV.  Several recent 
papers~\cite{Masiero:2004ft,Ullio:2001qk,Arvanitaki:2004df} have 
discussed the feasibility of detecting such a signal, for lower-scale AMSB 
scenarios at currently functioning or currently planned detectors.  Our approach 
will be rather to assume the specific, PeV-scale theory outlined above and 
assess what performance attributes would render a detector capable of 
observing the \(\gamma\)-ray signature of such particles.    
 
\indent
The cross-sections for the annihilation for a pair of Winos or Higgsinos 
into a pair of photons and into a photon and a Z in anomaly-mediated 
SUSY-breaking scenarios have been examined by several 
authors~\cite{Bergstrom:1997fh}.  In both the Wino and Higgsino 
cases, each cross-section tends toward an asymptotic value as 
\(m_{\mathrm{LSP}}\) increases.  For \(m_{\mathrm{LSP}}\) greater than a 
few hundred GeV, \(\sigma v(\gamma\gamma)\) and \(\sigma v(Z\gamma)\) may 
effectively be replaced with their asymptotic values, which for a Wino 
LSP are:   

\begin{eqnarray}
  \sigma_{\tilde{W}} v(\gamma\gamma) \simeq 4.0 \times 10^{-27} \mbox{ }
    \mathrm{cm}^{-3}\mathrm{s^{-1}}   \\
  \sigma_{\tilde{W}} v(Z\gamma) \simeq 9.0 \times 10^{-27} \mbox{ }
  \mathrm{cm}^{-3}\mathrm{s^{-1}},
\end{eqnarray}
and for a Higgsino LSP, are
\begin{eqnarray}
  \sigma_{\tilde{H}} v(\gamma\gamma) \simeq 9.0 \times 10^{-29} \mbox{ }
    \mathrm{cm}^{-3} \mathrm{s^{-1}}   \\
  \sigma_{\tilde{H}} v(Z\gamma) \simeq 2.0 \times 10^{-29} \mbox{ }
    \mathrm{cm}^{-3}\mathrm{s^{-1}}.
\end{eqnarray}
To be able to resolve the two lines would be an excellent test of the theory,
but to do so for a 2.3 TeV Wino would require an energy resolution 
\(\Delta E/E\) of better than 3.5\%.  Although future satellite facilities
comparable to GLAST may have this kind of energy resolution, these telescopes 
are unlikely to discover dark matter from PeV-scale split supersymmetry for 
reasons we shall soon make clear.  For ground-based detectors, which have far 
coarser energy resolutions (on the order of 10-20\%), the two signals will be 
indistinguishable and add together to form a single `line'.

\section{Halo Models and Dark Matter Distribution}

\indent

While the particle properties (mass, cross-section, etc.) of PeV-scale AMSB dark 
matter are more or less specified by the theory we have chosen, the distribution 
of that dark matter in the galaxy has yet to be specified.  The shape of this 
distribution is important in determining the requirements for indirect 
detection of a 2.3 TeV Wino, but is also not well known.  Thus, rather than 
choosing any particular model, we focus on a representative set of halo
profiles.  These profiles are derived from numerical 
simulations~\cite{Moore:1999gc,Navarro:1995iw}, and most take the form

\begin{equation}
  \label{eq:albega}
  \rho(r)=
  \frac{\rho_{0}}{(r/R)^{\gamma}(1+(r/R)^{\alpha})^{(\beta-\gamma)/\alpha}},
\end{equation}
where the three power-law indices \(\alpha\), \(\beta\), and \(\gamma\), along 
with the characteristic radius R, define a given model.  The models we 
examine here include the Moore et al. profile (a relatively cuspy model), 
a pair of isothermal models, one with a smooth density distribution, the other
with some clumping of the dark matter, and the widely used Navarro-Frenk-White 
profile.  The choice of \(\alpha\), \(\beta\), and \(\gamma\) which defines 
each model is given in 
table~\ref{tab:abgvals}.  In addition to these, we include in our 
analysis the halo profile proposed by Burkert et al.\ \cite{Burkert:1995yz}, 
for which the CDM density is modeled not by equation (\ref{eq:albega}), but
by

\begin{equation}
  \rho(r)=
  \frac{\rho_{0}r_{0}^{3}}{(r-r_{0})(r^{2}-r_{0}^{2})},
\end{equation}
where \(r_{0}\) is a fiducial distance parameter.  It has also been 
suggested that the presence of a massive black hole at the center of 
the galaxy could significantly alter the halo profile through the addition of 
a density spike at the galactic center~\cite{Gondolo:1999ef}.  However, as
there is some debate over the precise effect the black hole would have, we
will not consider this situation here.

\begin{figure}[ht!]
  \begin{center}
    \includegraphics{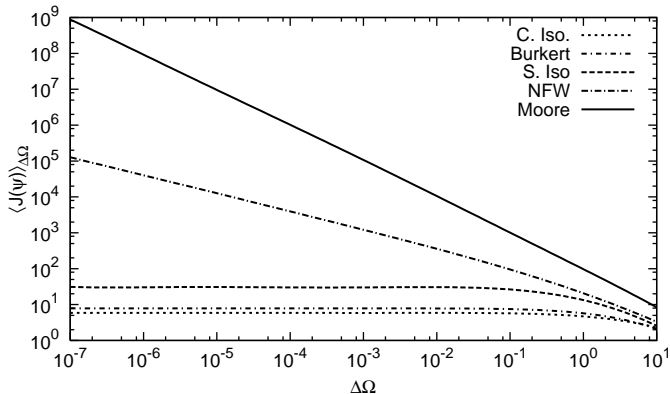}
  \end{center}
  \caption{\(\langle J(\psi)\rangle_{\Delta\Omega}\) (the line-of-sight integral 
    through the halo density squared, averaged over the angular acceptance 
    \(\Delta\Omega\)) vs. \(\Delta\Omega\) for 
    several halo profiles: smooth isothermal, clumpy isothermal, Burkert,  
    NFW, and Moore et al.\label{fig:jpsis}}
\end{figure}

\indent
For a given distribution of dark matter \(\rho(r,\psi)\), the observed 
integral flux of \(\gamma\)-rays (usually expressed in \(\mbox{cm}^{-2}s^{-1}\)) 
from LSP annihilations around the galactic center is

\begin{equation}
  \Phi=(\sigma v)n_{\gamma}\frac{1}{4\pi m_{\chi}}
  \int_{L}\rho^{2}(\psi,s)\mathit{ds},
\end{equation}
where \(\rho\), the CDM mass density, depends on both the line-of-sight
distance element \(s\), the number \(n_{\gamma}\) of photons produced
per decay, and the angle \(\psi\) away from the galactic center,
and the integral is evaluated along the line of sight.  It is common to
abstract the density integral, which depends on the halo profile but is 
independent of the particle physics, by defining the quantity

\begin{equation}
  J(\psi)\equiv\frac{1}{8.5\mbox{ kpc}}
  \left(\frac{1}{0.3\mbox{ GeV}}\right)^{2}\int_{L}\rho^{2}(\psi,s)ds.
\end{equation}   
For a detector with angular acceptance \(\Delta\Omega\), the relevant quantity
is not \(J(0)\), but rather \(\langle J(\psi)\rangle_{\Delta\Omega}\), the 
average of
\(J(\psi)\) over the solid angle given by \(\Delta\Omega\).  This quantity
depends only on the value of \(\Delta\Omega\) and the choice of halo profile.
In figure~\ref{fig:jpsis}, we show the relationship between
\(\langle J(\psi)\rangle_{\Delta\Omega}\) and \(\Delta\Omega\) for some of the 
most commonly used halo models.

\begin{table}[t!]
  \begin{center}
    \begin{tabular}{|l|cccc|} \hline
    & \(\alpha\) & \(\beta\) & \(\gamma\) & \(R\) \\ \hline
    Isothermal profiles & 2.0 & 2.0 & 0 & 3.5 \\
    NFW & 1.0 & 3.0 & 1.0 & 20.0 \\
    Moore et al. & 1.5 & 3.5 & 1.5 & 28.0 \\ \hline
    \end{tabular}
    \end{center}
  \caption{The defining parameters \(\alpha\), \(\beta\), \(\gamma\) and \(R\) 
    (see equation \ref{eq:albega}) for the halo models we examine.  \(R\) is given in 
    kpc.\label{tab:abgvals}}
\end{table}

\section{Detection and Instrumental Limitations}

\indent

\begin{figure}[ht!]
  \begin{center}
    \includegraphics{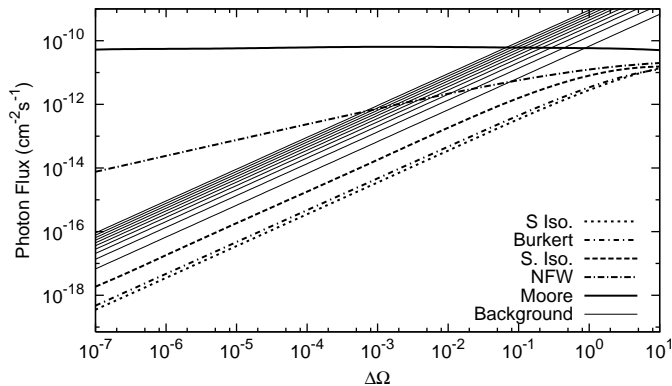}
  \end{center}
  \caption{The expected flux from the annihilation of 2.3 TeV Winos, as a 
    function of \(\Delta\Omega\), that would be detected by a 
    \textit{satellite detector} aimed at the galactic center.  Also included 
    is the 
    anticipated background flux at such a detector for different values of
    detector
    energy resolution, ranging from \(\Delta E/E = 50\%\) (top line) to
    \(\Delta E/E = 5\%\) (bottom line).  It 
    should be noted that the spread in the \textit{signal} is
    smaller than the width of the energy bin.  From this, it is apparent that 
    for the NFW and Moore
    et al.\ profiles, the prospects for detection increase with better angular
    resolution (decreasing 
    \(\Delta\Omega\)).  For a 1.1 TeV Higgsino LSP, the resulting curves are 
    similar, but the signal is two orders of magnitude lower, and the 
    background flux is increased by a factor of \(\sim10\).\label{fig:fluSp}}
\end{figure}

\begin{figure}[ht!]
  \begin{center}
    \includegraphics{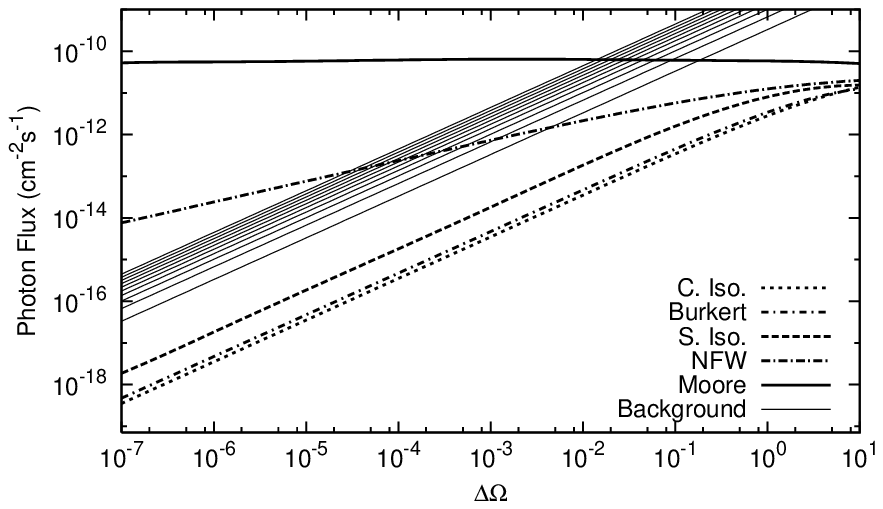}
  \end{center}
  \caption{The expected flux from the annihilation of 2.3 TeV Winos, as a 
    function of \(\Delta\Omega\), that would be detected by a 
    \textit{Cherenkhov detector} aimed at the galactic center.  Also included 
    is the 
    anticipated background flux at such a detector for different values of 
    detector energy resolution, ranging from \(\Delta E/E = 50\%\) (top line) to
    \(\Delta E/E = 5\%\) (bottom line).  
    It should be noted that the spread in the \textit{signal} is
    smaller than the width of the energy bin.  From this, it is apparent that 
    for the NFW and Moore
    et al. profiles, the prospects for detection increase with better angular
    resolution (decreasing 
    \(\Delta\Omega\)).  For a 1.1 TeV Higgsino LSP, the resulting curves are 
    similar, but the signal is two orders of magnitude lower, and the 
    background flux is increased by a factor of \(\sim10\).\label{fig:fluCh}}
\end{figure}

We now turn to a discussion of the properties of \(\gamma\)-ray detectors
and the physical limitations to which they are subject. 
The relevant quantities, in terms of telescope performance, are effective
area, energy resolution, angular resolution, and field of view.  Significant  
improvements in energy resolution could effectively cut the background in 
the energy range of interest by an order of magnitude (see 
figures~\ref{fig:fluSp} 
and~\ref{fig:fluCh}), while better angular resolution can improve
the signal-to-background ratio, depending on the halo profile.  For a cuspy
profile, such as the Moore et al.\ or NFW profile, 
\(\langle J(\psi)\rangle_{\Delta\Omega}\)
increases rapidly with decreasing \(\Delta\Omega\) and minimizing the
angular acceptance enhances the visibility of the signal.  For a less cuspy
profile, the signal-to-background ratio decreases as \(\Delta\Omega\) becomes
smaller and the observation strategy would then be to relax the angular 
acceptance as much as possible within the detector's field of view.

\indent
Satellite detectors, such as the soon-to-be-launched pair-production
telescope GLAST, can have excellent angular resolution and energy resolution.  
The field of view for GLAST will be on the order of a steradian, its 
angular resolution in the TeV range will be \(\sim0.1^{\circ}\), and its
energy resolution \(\Delta E/ E\) on the order of 4\%.  Despite all 
this, all space-based telescopes are limited by collection area constraints 
(\(A_{\mathrm{eff}}\) tends to be on the order of 
\(10^{4}\mbox{ }\mathrm{cm}^{2}\)), which will make it difficult for
any space telescope to detect a large enough number photons in the TeV range 
to register a \(5\sigma\) discovery. 

\indent     
In contrast, atmospheric Cherenkhov detectors (ACTs) such as the HESS array, 
which operate by observing showers of Cherenkhov light that result when 
high-energy \(\gamma\)-rays enter the Earth's upper atmosphere, can have
far larger effective areas.  The area of the Cherenkhov light pool on the 
ground is \(\sim5\times10^{8}\mbox{ }\mathrm{cm}^{2}\), 
which defines the scale of the detector's collection area, which is 
comparatively enormous.  However, 
uncertainties in reconstructing the energy of the primary photon from the 
properties of the radiation shower place limits on the energy resolution (a 
single imaging detector can achieve \(\Delta E/E\simeq 30 -40\%\); an array
of parallel detectors, \(10-15\%\)).  Still, because of their large collection
area, ACTs are probably the best bet for the detection of Wino or
Higgsino dark matter from PeV-scale split supersymmetry.      
     
\indent
In addition to the performance attributes discussed above, space-based and
ground-based telescopes also `see' different backgrounds.  For a satellite 
detector such as GLAST, the background is the actual diffuse 
gamma-ray background, which for energies in the TeV range is less well-known.  
The best that can currently be done is to make the assumption one can extrapolate
the power law spectrum from EGRET data (good up to \(\sim100 \mbox{ GeV}\)) to 
higher energies~\cite{Hunger:1997we,Sreekumar:1997un,Strong:2004de}, which yields 
a power-law of the form    
 
\begin{equation}
  \label{eq:alphaSpace}
  \frac{dn_{\mathrm{BG}}}{d\Omega dE}=N_{0}
  \left(\frac{E}{1\mbox{ GeV}}\right)^{-\alpha}\mathrm{cm}^{-2}\mathrm{s}^{-1}
  \mathrm{GeV}^{-1}\mathrm{sr}^{-1},
\end{equation}
with a numerical prefactor \(N_{0}\) on the order of \(10^{-6} \mbox{ cm}^{2}\)
and an exponent \(\alpha\) somewhere between 2.0 and 2.5.  Of course GLAST,
when it turns on, will also provide a great deal of information on the 
diffuse \(\gamma\)-ray background at high energies, which will dramatically 
reduce the uncertainty in \(\alpha\) and \(N_{0}\).    
\begin{table}[t!]
  \begin{center}
    \begin{tabular}{|l|ccccc|} \hline
      & \(A_{\mathrm{eff}}\) & \(\Delta E/E\) & \(\Delta\Omega_{\mathrm{min}}\)
      & \(\epsilon_{\mathrm{had}}\) & \\ \hline
      WHIPPLE (Arizona) & $3.5 \times 10^{8}$ & 30\% & $1.88 \times 10^{-5}$ & 1.0 & \\
      GRANITE II (Arizona)  & $5\times 10^{8}$ & 20\% & $9.56 \times 10^{-6}$ & 1.0 & \\
      HESS (Namibia) & $7 \times 10^{8}$ & 15\% & $9.56 \times 10^{-6}$ & 0.25 & \\
      VERITAS (Arizona) & $1 \times 10^{9}$ & 15\% & $3.83 \times 10^{-7}$ & 0.25 & \\
      EGRET (Satellite) & $1 \times 10^{4}$ & 15\% & $3.22 \times 10^{-2}$ & - & \\
      GLAST (Satellite) & $1.5 \times 10^{4}$ & 4\% & $9.56 \times 10^{-6}$ & - & \\ \hline
      Generic ACT & $1.5 \times 10^{9}$ & 10\% & $1.00 \times 10^{7}$ & 
      0.25 & \\
      Generic PPT & $2 \times 10^{4}$ & 1\% & $1.00 \times 10^{-7}$ & - & \\
      \hline
    \end{tabular}
  \end{center}
  \caption{The performance parameters for current and planned \(\gamma\)-ray 
    telescopes, including both ACTs (WHIPPLE, GRANITE II, HESS, and VERITAS) 
    and space telescopes (EGRET and GLAST).  Also included are the parameters
    used for the generic atmospheric Cerenkhov telescope (ACT) and space-based pair 
    production telescope (PPT) we have used in our analysis.\label{tab:detectors}} 
\end{table} 

\indent
The background for atmospheric Cherenkhov detectors consists mainly of cascade 
events triggered by cosmic-ray protons, electrons, etc., which dominate (by 
an order or two of magnitude) over the diffuse gamma-ray background.  Showers
initiated by leptons (predominately electrons) are indistinguishable from 
gamma-ray cascades, whereas hadronic showers can be differentiated to a degree 
due to the cascade's shape and to the time spread of the light pulse.  While 
these backgrounds are higher than those seen by satellite detectors, their 
spectra are reliably known up to 5 TeV .  The power-law 
behavior~\cite{Bergstrom:1997fj} for each is given below:          

\begin{eqnarray}
  \label{eq:alphaACT}
  \frac{dN_{\mathrm{had}}}{dEd\Omega}=1.0\cdot 10^{-2}\epsilon_{\mathrm{had}}
    \left(\frac{E_{0}}{1\mbox{ GeV}}\right)^{-2.7}\mbox{ }
    \mathrm{cm}^{-2}\mbox{s}^{-1}\mbox{ GeV}^{-1}\mathrm{sr}^{-1} \\
  \frac{dN_{\mathrm{e^{-}}}}{dEd\Omega}=6.9\cdot 10^{-2} 
    \left(\frac{E_{0}}{1\mbox{ GeV}}\right)^{-3.3}\mbox{ }
    \mathrm{cm}^{-2}\mbox{s}^{-1}\mbox{ GeV}^{-1}\mathrm{sr}^{-1},   
\end{eqnarray}
where we have replaced \(N_{0}\) and the power-law index \(\alpha\) with their
explicit numerical values. 
The factor of \(\epsilon_{\mathrm{had}}\) has been included to account for 
improved hadronic rejection techniques (the default value of 
\(\epsilon_{\mathrm{had}}\)=1 corresponds to the Whipple telescope:
instruments such as HESS and VERITAS have already improved on this by 
a factor of four).  

\indent
In figures~\ref{fig:fluSp} 
and~\ref{fig:fluCh}, we show the 
expected gamma-ray flux from
the 2.3 TeV line as a function of angular acceptance, along with the
expected background flux at a generic ACT and satellite detector
whose performance attributes are slightly better than those of any
currently planned facility (see
table~\ref{tab:detectors} for performance specifics) for a variety of 
different energy resolutions ranging from 5\% to 50\%.  It is
apparent that the integral background flux seen by a satellite detector 
will generally be at least two orders of magnitude below that seen by an ACT, 
regardless of the precise power-law form of the diffuse \(\gamma\)-ray 
background.
   
\begin{figure}[ht!]
  \begin{center}
    \includegraphics{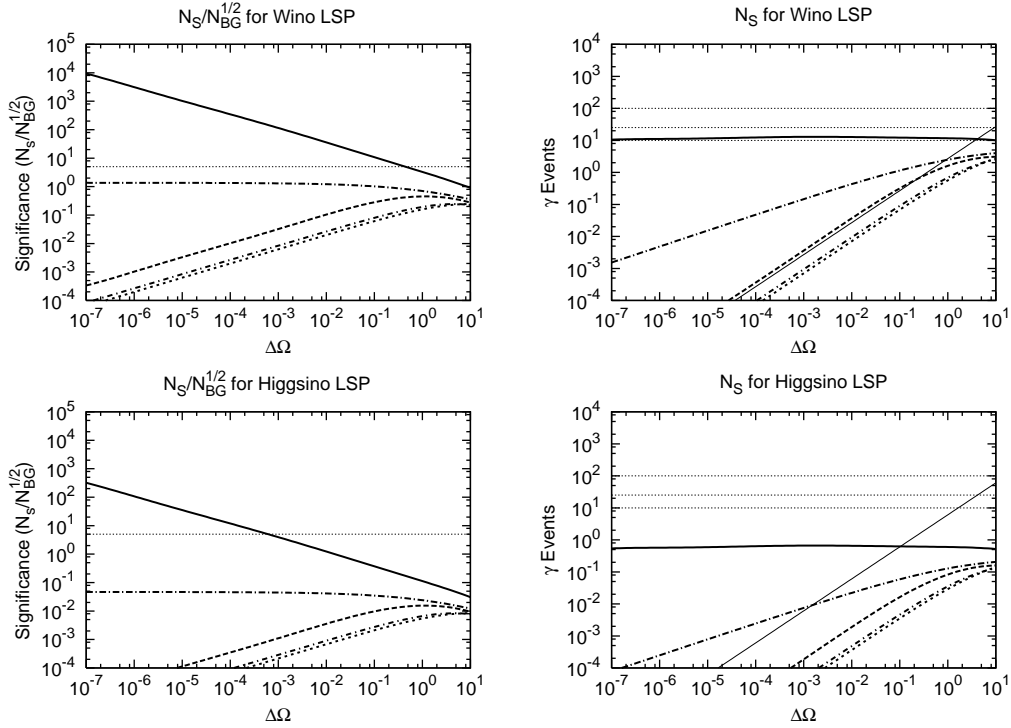}
  \end{center}
  \caption{The ratio of \(N_{\mathrm{signal}}/\sqrt{N_{\mathrm{background}}}\)  
    (left panels) and total number of photons (right panels) collected by a generic 
    \textit{space telescope} with an effective area of 
    \(2 \times 10^{4}\mbox{ cm}^{2}\) 
    and an energy resolution of 1\%, over a range of \(\Delta\Omega\), after 
    \(10^{7}\mbox{ s}\) (about 1/3 of an active year) of viewing time, and for both 
    2.3 TeV Wino (top panels) 
    and 1.1 TeV Higgsino LSP (bottom panels).  See figure~\ref{fig:fluCh} 
    caption for the halo model key.  The 
    threshold for \(5\sigma\) discovery has been 
    included for reference in the significance graphs, and contours corresponding
    to 10, 25, and 100 events have been included in the event
    count graphs.  It can be seen here that for such a space telescope, no halo 
    profile is capable of producing the
    25 events necessary for detection, and that only for the Moore et al.\ profile is 
    the significance criterion even achieved.\label{fig:EventsSigSp}}
\end{figure}

\indent
In order for the signal registered at any detector to be interpreted as a 
discovery, two conditions must be met: first, the significance level (the
ratio of \(N_{S}\), the total number of signal photons registered, to 
\(\sqrt{N_{\mathrm{BG}}}\), where \(N_{\mathrm{BG}}\) is the total number of
background photons registered) must exceed \(5\sigma\); second, the total number
of detected photons must exceed 25, the threshold below which Poisson statistics
give an equivalent confidence limit\footnote{While the likelihood of random 
statistical fluctuations at 
the \(5\sigma\) level increases with improved energy resolution, these can be 
differentiated from a true signal by requiring the signal to be consistent over
multiple trials.}.  These requirements, when written explicitly
in terms of \(A_{\mathrm{eff}}\), \(\Delta E/E\), \(\Delta\Omega\), and 
observation time, are

\begin{equation}
  \label{eq:fivesigma}
  (.68)^{2}\left(\frac{\Phi(\Delta\Omega)\sqrt{A_{\mathrm{eff}}t}}{\sqrt{\Phi_{\mathrm{BG}}(\Delta\Omega,\Delta E/E, \epsilon_{\mathrm{had}})}}\right)\geq 5
\end{equation}
\begin{equation}
  \label{eq:eventcount}
  \Phi(\Delta\Omega)A_{\mathrm{eff}}t\geq 25
\end{equation}    

\begin{figure}[ht!]  
  \begin{center}
    \includegraphics{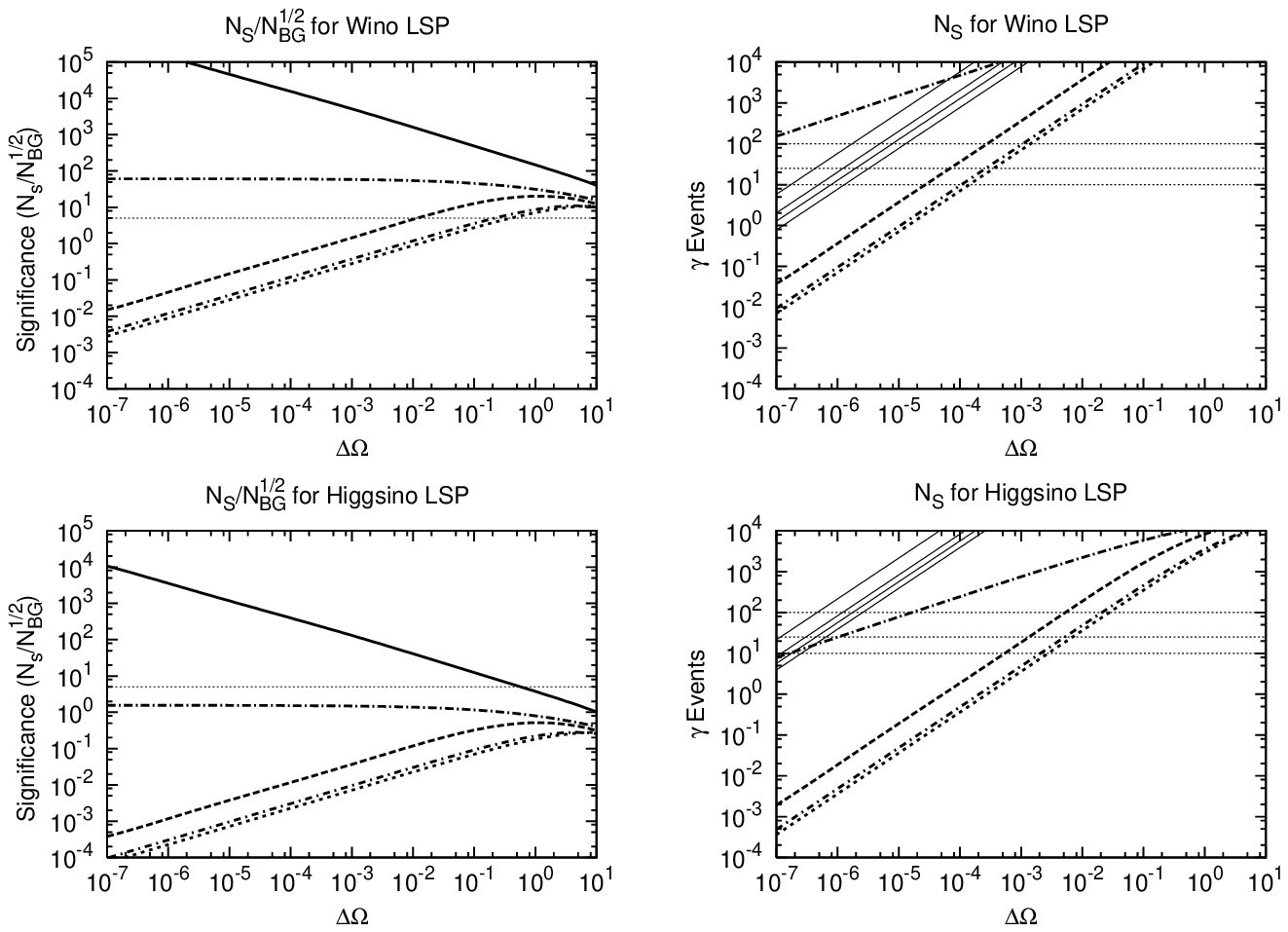}
  \end{center}
  \caption{The ratio of \(N_{\mathrm{signal}}/\sqrt{N_{\mathrm{background}}}\)  
    (left panels) and total number of photons (right panels) collected by a generic 
    \textit{Cherenkhov array} with an effective area of 
    \(1.5\times 10^{9}\mbox{ cm}^{2}\)
    and an energy resolution of 10\%, over a range of \(\Delta\Omega\), after  
    \(10^{7}\mbox{ s}\) (about 1/3 of an active year) of viewing time, and for 
    both 2.3 TeV Wino (top panels) 
    and 1.1 TeV Higgsino LSP (bottom panels).  See figure~\ref{fig:fluCh}
    caption for the halo model key.  The 
    threshold for \(5\sigma\) discovery has been 
    included for reference in the significance graphs, and contours corresponding
    to 10, 25, and 100 events have been included in the event
    count graphs.  It can be seen here that there are real prospects for detection 
    with such a Cherenkhov detector, provided that the galactic CDM halo density
    resembles the NFW or Moore et al.\ profiles.\label{fig:EventsSigCh}}
\end{figure}

Because the total number of collected photons is directly proportional to the 
collecting area, the second criterion implies that an ACT, with a characteristic 
\(A_{\mathrm{eff}}\) on the order of \(10^{8} - 10^{9} \mbox{ cm}^{2}\), will be
a more useful for detecting a PeV-scale dark matter signal than a satellite
detector with an \(A_{\mathrm{eff}}\) of \(\sim10^{4}\mbox{ cm}^{2}\), despite
the smaller background seen by the latter.  In figures~\ref{fig:EventsSigSp} 
and~\ref{fig:EventsSigCh}, we plot the the total number of signal photons 
recorded by our generic satellite detector and Cherenkhov array for a range 
of \(\Delta\Omega\) and an exposure time of \(10^{7}\) s, in both the Wino and 
Higgsino LSP cases.  Even with the liberal assumptions of a Wino-like LSP and a 
galactic halo density described by the Moore et al. profile, our telescope (and
hence also GLAST) would not register a sufficient number of photons to signal a 
discovery.

\indent
The situation is far more hopeful for Cherenkhov detectors.  Since the field
of view an ACT can attain is on the order of \(10^{-3} \mbox{ cm}^{2}\),
detection would still be out of reach if the dark matter distribution was
less sharply cusped (such as in the isothermal and Burkert profiles), but
would be possible for a more concentrated dark matter halo.  It is an interesting
coincidence that the NFW profile nearly demarcates the line between detection 
and non-detection for presently operational facilities: if the actual dark 
matter distribution is cuspier than that given by the NFW profile, 
the \(\gamma\)-ray signature of Wino dark matter in PeV-scale split 
supersymmetry 
should be detectable at the next generation of ACTs; if the actual profile 
is much less sharply peaked (e.g. if it resembles the smooth isothermal case), 
it is unlikely that such a signal would ever be detectable at an ACT.  It should 
also be noted that a 1.1 TeV Higgsino LSP would be more difficult to detect 
than a 2.3 TeV Wino (only for the Moore et al. profile is the \(5\sigma\) 
requirement from equation (\ref{eq:fivesigma}) satisfied), both because the 
annihilation cross-section into 
\(\gamma\)-rays is smaller and because the background is around a factor of ten 
higher\footnote{This follows simply because the background photon spectrum obeys 
a power-law.
Although the power-law index \(\alpha\) is slightly different for ACTs and 
satellite detectors (see equation (\ref{eq:alphaACT}) and discussion below 
(\ref{eq:alphaSpace})), the results for both are comparable.} at 1.1 TeV than 
at 2.3 TeV.  

\section{Future Detectors}

\begin{figure}[ht!]
  \centering
    \includegraphics{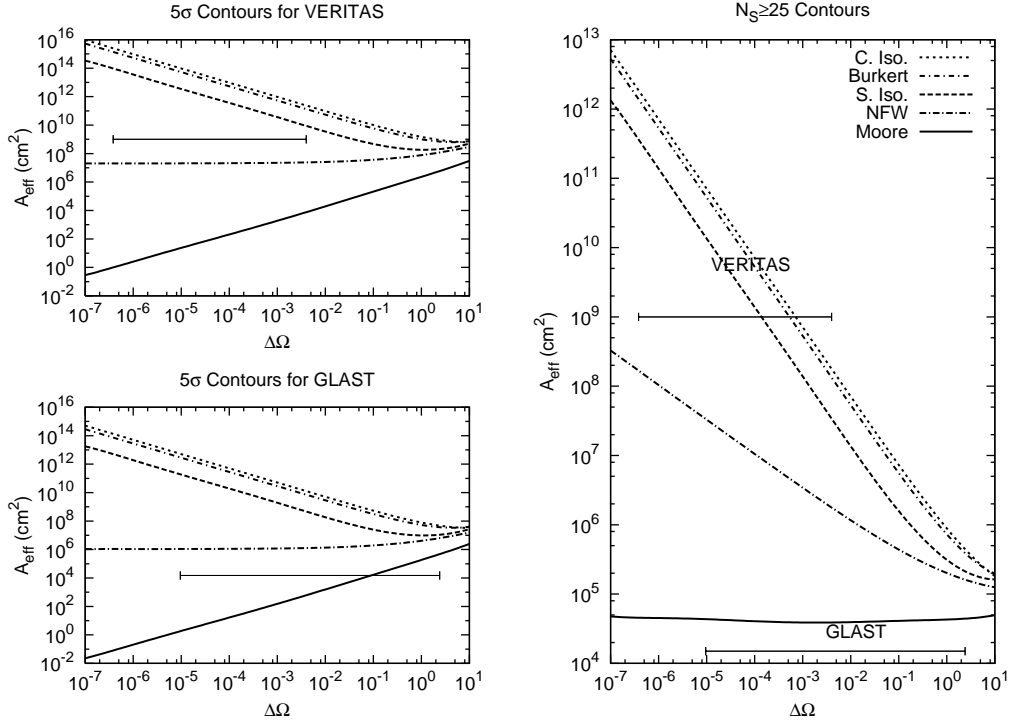}
  \caption{Detection boundary contours in \(A_{\mathrm{eff}}\)-\(\Delta\Omega\) 
    parameter space for the \(\gamma\)-ray signature of a 2.3 TeV Wino,
    based on the \(5\sigma\) significance requirement from 
    equation~\ref{eq:fivesigma} for VERITAS (top left panel), with 
    \(A_{\mathrm{eff}}=1\times 10^{9}\mbox{ cm}^{2}\) and \(\Delta E/E\), and 
    GLAST (bottom left panel), with 
    \(A_{\mathrm{eff}}=1.5\times 10^{4}\mbox{ cm}^{2}\);
    and based on the \(N_{S}\geq25\) event count requirement from 
    equation~\ref{eq:eventcount} (right panel), for a variety of halo profiles.  Bars 
    showing the range of angular acceptances that can be chosen at VERITAS and 
    GLAST have 
    also been included.  In order to register a discovery at either of these
    facilities for a given halo model, \textit{both} the \(5\sigma\) and 
    \(N_{s}\geq25\) 
    contours for that model must lie below the bar corresponding to that facility. 
    \label{fig:ContCh}}
\end{figure}

\indent
An interesting question to ask, especially if current facilities would be unable
to register a discovery, is what attributes would a detector need in order to
conclusively discover Wino or Higgsino dark matter in a PeV-scale split 
supersymmetry scenario.  Since collection area and angular acceptance (which
may be adjusted as desired over a window ranging from a detector's
angular resolution (in steradians) to its field of view) have the most 
significant effect on the detection capabilities of any given instrument, we
will focus on the effect of improvements in \(A_{\mathrm{eff}}\) and 
\(\Delta\Omega\).  In figure~\ref{fig:ContCh}, we show the contours defined
by equations (\ref{eq:fivesigma}) and (\ref{eq:eventcount}) in 
\(A_{\mathrm{eff}}\)-\(\Delta\Omega\) space for \(10^{7}\mbox{ s}\) of viewing
time, where the \(\Delta E/E\) values used in computing the significance limits
are those for the GLAST telescope and the VERITAS array (see 
table~\ref{tab:detectors}), along with the attributes of GLAST and VERITAS 
themselves, which appear as bars due to our freedom to adjust \(\Delta\Omega\).
The criterion for discovery at either of these facilities is that for a given halo 
model, both the \(5\sigma\) and \(N_{s}\geq25\) contours for that model lie below 
the corresponding bar.  From this figure, it is evident that GLAST, primarily
due to its small effective area, would be unable to detect CDM from PeV-scale
split supersymmetry at all.  Detection would be possible at VERITAS, 
provided that the dark matter distribution isn't too much flatter than the NFW 
profile.  If the halo resembles the Burkert or isothermal profiles, however,
neither facility (and no telescope present or currently planned) would be able to 
register a \(5\sigma\) discovery.

\indent
Because of their inherently large collection areas, Cherenkhov detectors
are currently the best bet for the discovery of dark matter from PeV-scale
split supersymmetry.  The collection area of an ACT may be increased by the 
addition of more telescopes in the array, but the increment supplied by each 
individual telescope is merely additive, meaning it would not be feasible to 
raise \(A_{\mathrm{eff}}\) by two orders of magnitude over that of VERITAS.  It
would, however, be possible to engineer an ACT with a larger field of view:
the fields of view for present and planned facilities are kept on the order
of \(10^{-3}\mbox{ sr}\) primarily as a mechanism to deal with the effects
of secondary electron scattering on the detector's low-energy response which
will not affect observations in the TeV range.  Operating at a large 
\(\Delta\Omega\), such an instrument would be able to register a discovery for
a range of more gently sloping halo profiles. 

\indent
Since ACTs are intrinsically limited in energy resolution to 
\(\Delta\Omega\sim10\%\) by uncertainties in reconstructing the energy of the
initial photon from its cascade products~\cite{Weekes:2003}, future facilities 
would still be 
unable to see the telltale indication that an observed \(\gamma\)-ray signal
is the result of dark matter annihilation: 
the resolution of the \(Z\gamma\) and \(\gamma\gamma\) lines.  Satellite 
detectors are already approaching this level of energy resolution, and so it 
is of particular interest whether a space-based facility could ever be built
that would be able to detect them\footnote{Another advantage satellite
detectors have over their ground-based counterparts is that their large fields 
of view could in principle allow for detection in even the most conservative 
of halo models.}.  The difficulty, of course, is that even count
considerations would mandate that the detection array be at least 
\(100\mbox{ m}\) on a side, and that keeping such an object operational for
\(10^{7}\mbox{ s}\) (considering potential disasters relating to solar wind, 
space debris, etc. that such an object would face), let alone launching it into 
into space, amount to a nearly insurmountable problem.  It might be possible
someday to construct such a detection array on the far side of the moon, where
the weight and area of the detector would not be an impediment, but unless a
project of this sort were undertaken, it is unlikely that we would ever see
the conclusive double-line signal of dark matter in PeV-scale split 
supersymmetry.  

\section{Acknowledgements}
I would like to thank Tim McKay, Ting Wang, and James Wells for useful 
discussions.

\end{document}